\PassOptionsToPackage{unicode}{hyperref}
\PassOptionsToPackage{hyphens}{url}
\documentclass[aps,prl,reprint,superscriptaddress]{revtex4-2}
\usepackage{xcolor}
\usepackage{amsmath,amssymb}
\usepackage{iftex}
\ifPDFTeX
  \usepackage[T1]{fontenc}
  \usepackage[utf8]{inputenc}
  \usepackage{textcomp} 
\else 
  \usepackage{unicode-math} 
  \defaultfontfeatures{Scale=MatchLowercase}
  \defaultfontfeatures[\rmfamily]{Ligatures=TeX,Scale=1}
\fi
\usepackage{lmodern}
\ifPDFTeX\else
\fi
\IfFileExists{upquote.sty}{\usepackage{upquote}}{}
\IfFileExists{microtype.sty}{
  \usepackage[]{microtype}
  \UseMicrotypeSet[protrusion]{basicmath} 
}{}
\makeatletter
\@ifundefined{KOMAClassName}{
  \IfFileExists{parskip.sty}{%
    \usepackage{parskip}
  }{
    \setlength{\parindent}{0pt}
    \setlength{\parskip}{6pt plus 2pt minus 1pt}}
}{
  \KOMAoptions{parskip=half}}
\makeatother
\usepackage{graphicx}
\makeatletter
\newsavebox\pandoc@box
\newcommand*\pandocbounded[1]{
  \sbox\pandoc@box{#1}%
  \Gscale@div\@tempa{\textheight}{\dimexpr\ht\pandoc@box+\dp\pandoc@box\relax}%
  \Gscale@div\@tempb{\linewidth}{\wd\pandoc@box}%
  \ifdim\@tempb\p@<\@tempa\p@\let\@tempa\@tempb\fi
  \ifdim\@tempa\p@<\p@\scalebox{\@tempa}{\usebox\pandoc@box}%
  \else\usebox{\pandoc@box}%
  \fi%
}
\def\fps@figure{htbp}
\makeatother
\ifLuaTeX
  \usepackage{luacolor}
  \usepackage[soul]{lua-ul}
\else
  \usepackage{soul}
\fi
\usepackage{bookmark}
\IfFileExists{xurl.sty}{\usepackage{xurl}}{} 
\hypersetup{
  hidelinks,
  pdfcreator={LaTeX via pandoc}}

\date{\today}
\begin{document}

\title{Passive Imaging with Quantum Advantage}

\author{Li Gong}
\affiliation{Department of Physics, University of Oxford, Oxford, OX1 3PU, UK}
\affiliation{Quantum Innovation Centre (Q.InC), Agency for Science Technology and Research (A*STAR), Singapore}

\author{Aonan Zhang}
\affiliation{Department of Physics, University of Oxford, Oxford, OX1 3PU, UK}

\author{Madhura Ghosh Dastidar}
\affiliation{Department of Physics, University of Oxford, Oxford, OX1 3PU, UK}

\author{Alexander Duplinskii}
\affiliation{Department of Physics, University of Oxford, Oxford, OX1 3PU, UK}

\author{A. I. Lvovsky}
\affiliation{Department of Physics, University of Oxford, Oxford, OX1 3PU, UK}

\begin{abstract}
Far-field optical imaging inevitably involves low-pass spatial filtering, limiting the resolution. Moreover, conventional imaging suppresses high spatial frequency components close to the cutoff, making them invisible under noise --- particularly the shot noise arising from discrete and random nature of quantum light. Here we propose and implement a method for reducing the effect of this noise by optically pre-processing  the incoming light prior to detection, thereby optimizing the quantum measurement performed on it. Our scheme, termed Fourier Domain Division (FDD), partitions the Fourier plane into multiple regions for independent detection and subsequent post-processing for image reconstruction. By analyzing the quantum and classical Fisher information, we show that our method is advantageous with respect to direct imaging for high spatial-frequency components. As a result, the number of photons required to achieve a certain signal-to-noise-ratio in the Fourier domain is reduced, thus enhancing the overall resolution in the photon-starved regime. We demonstrate our method in microscopy, achieving 5-fold improvement of Fisher information on high spatial-frequency components. 
Unlike active super-resolution methods, FDD is passive, making it broadly applicable in microscopy and other imaging scenarios where active illumination is impractical, including astronomy and remote sensing. Our work establishes a general strategy for designing quantum optimized superresolution imaging systems, bridging fundamental quantum limits, practical image analysis and computer vision applications.
\end{abstract}

\maketitle

\section{Introduction}

In conventional optical microscopy, spatial resolution is fundamentally constrained by diffraction, a direct consequence of the wave nature of light. Abbe interpreted this limit as a restriction in the Fourier domain of the image field imposed by the aperture of the objective lens \cite{Ref1}. For incoherent intensity imaging, this results in the optical transfer function (OTF) of a nearly conical shape, meaning that the image is distorted in favor of low spatial frequency components, resulting in a further loss of resolution \cite{Ref24}. To compensate for this loss, various deconvolution techniques are used. However, the capability of deconvolution  rapidly degrades in the presence of noise \cite{Ref25}, which can be both of fundamental and technical nature. 

Here we focus on the former case --- namely the shot noise, which results from the discrete and random nature of quantum light, and becomes dominant under the conditions of low intensity. We propose to address this issue by treating it as a quantum sensing problem. That is, we consider the  quantum state of the optical field that passes through the objective lens and look for the measurement of this state that enhances the image reconstruction precision. 

This approach enables us to use the  tools from quantum estimation theory to quantify the measurement precision. Central to this framework is the notion of parameter --- a quantitative feature of the quantum state that needs to be estimated. For a given set of parameters, the quantum Fisher information (QFI) sets the quantum Cramér-Rao bound (QCRB) --- the lowest possible variance for an unbiased estimator permitted by quantum mechanics, minimized over all possible measurements \cite{Ref6}. When the measurement is fixed, the achievable precision is governed by the classical Fisher information (FI) and its associated classical Cramér-Rao bound (CRB), which can never surpass the QCRB \cite{Ref6}.

The traditional diffraction limit has been re-examined from this quantum perspective \cite{Ref7}. A striking result has been that the QFI for estimating the separation between two incoherent point sources remains constant, independent of how small that separation is. The FI attainable with conventional direct imaging (DI) falls far below the QFI. On the other hand, the quantum limit can be saturated by measuring the optical field in the Hermite-Gaussian (HG) spatial mode basis \cite{Ref7}. This strategy has been explored extensively in theory \cite{Ref8,Ref9,Ref10} and demonstrated in a series of experiments \cite{Ref11,Ref12,Ref13,Ref14,Ref15}. 

Imaging, however, is a much more complex task than point source separation. The complexity is due to not only a much larger parameter set, but also correlations amongst the parameters and incompatibility between optimal measurements. HG basis measurement was proposed for imaging arbitrary two-dimensional objects \cite{Ref16}, and shown to offer resolution advantage, albeit outside the quantum domain \cite{Ref17}. Experimental validations now include coherent \cite{Ref18} and incoherent \cite{Ref19} imaging with wide-field illumination, as well as coherent imaging with a scanning, tightly focused beam \cite{Ref20}. In the quantum domain, however, the advantage of HG imaging has not been established. More generally, imaging as a quantum sensing problem is not well-understood: the ultimate quantum bound for general imaging scenarios has not yet been analyzed nor even a functional parameter set for this task has been established. 

This gap motivates our work on the quantum limit of imaging arbitrary samples and the design of optimal measurements. In our approach, we take the Fourier components of the sample intensity distribution as the parameters to be estimated, since accurately measuring high spatial-frequency information is closely related to the resolution. We determine the QFI for this parametrization: analytical calculations using reasonable approximations, validated by  numerical simulations, provide insight into the design of optimized measurements. Accordingly, we propose a measurement scheme called Fourier Domain Division (FDD), in which the Fourier plane of the imaging system is divided into multiple regions followed by intensity distribution measurement in the conjugate plane of the object. We develop an estimator to reconstruct the image from these measurements; our estimator is a generalized Wiener filter that effectively combines the information from all regions. Finally, we demonstrate FDD experimentally using a fluorescent resolution target, achieving a 1.1-fold improvement in resolution, and requiring 5-fold fewer photons to reach the same estimation accuracy for high spatial-frequency information. 

The innovation of our work consists not only in developing a specific imaging method with quantum advantage. Converting the complicated quantum optimization task to a relatively simple OTF engineering problem establishes a general strategy for constructing quantum optimized imaging methods. This strategy can incorporate resolution enhancement tools such as structured illumination or two-photon fluorescence. By bridging quantum estimation theory and practical imaging, our approach establishes a pathway to fully exploit the quantum nature of light for super-resolution imaging of arbitrary samples.


\section{Results}

\subsection{Quantum estimation theory for arbitrary sample}

An imaging system can be described as a low-pass filter in the Fourier domain, with the high-frequency components restricted by the objective lens aperture. The resolution is determined by the transmission and measurement precision of the high spatial frequency components.

To analyse the fundamental limits of an imaging system within the framework of quantum estimation theory, we first need a proper set of parameters that captures all resolvable structural information of the sample. Let the incoherent emission intensity distribution of the object (ground truth) be \(f(\mathbf{r})\). Its Fourier expansion is

\begin{equation}
f\left( \mathbf{r} \right) = a_{0} + \sum_{\substack{|\mathbf{k}|\leq k_c\\ k_x,k_y > 0}}^{}{a_{\mathbf{k}}\cos\left( \mathbf{k \cdot r} \right) + b_{\mathbf{k}}\sin\left( \mathbf{k \cdot r} \right)},
\label{eq:sample}
\end{equation}

We choose the amplitudes of the spatial Fourier components of the object \(a_{\mathbf{k}}\) and \(b_{\mathbf{k}}\) as the parameters to be estimated. \(f(\mathbf{r})\) is assumed normalized such that \(\iint_{}^{}{f(\mathbf{r})\ d^{2}\mathbf{r}} = 1\). The spatial frequency \(\mathbf{k}\) is truncated by the cutoff frequency \(k_{c} = \frac{4\pi}{\lambda}{\rm NA}\) of the incoherent imaging system under wide field illumination, where \(\lambda\) is the wavelength and NA is the numerical aperture \cite{Ref24}. We assume the object to be of finite width so the Fourier series, rather than a continuous Fourier transformation, can be applied. This assumption simplifies the calculations, but is not critical for the practical implementation of the method.

The quantum state of light in the image plane in each temporal mode is assumed to be in a weak thermal state \cite{Ref7,Ref21}, which can be modelled as \(\hat\rho = (1 - \delta)\left. \ |0 \right\rangle\left\langle 0| \right.\  + \delta\hat\rho_{1}\), where \(\delta\) is a small constant proportional to the emission rate and \(\hat \rho_{1}\) is the one-photon state,

\begin{equation}
\hat\rho_{1} = \iint_{}^{}{f\left( \mathbf{r} \right)e^{- i\mathbf{\hat K \cdot r}}\left. \ |\Psi \right\rangle\left\langle \Psi| \right.\ e^{ i\mathbf{\hat K \cdot r}}d^{2}\mathbf{r}}.
\label{eq:density_matrix}
\end{equation}
In this equation, \(\left. \ |\Psi \right\rangle = \iint_{}^{}{\Psi\left( \mathbf{r} \right)\left. \ |\mathbf{r} \right\rangle d^{2}\mathbf{r}}\), \(\Psi\left( \mathbf{r} \right)\) being the normalized amplitude point spread function (APSF) of the imaging system (\(\left\langle \Psi \middle| \Psi \right\rangle = 1\)), \(\left. \ |\mathbf{r} \right\rangle\) the photon position eigenket, and \(\mathbf{\hat K}\) the momentum operator, so that \(e^{- i\mathbf{\hat K \cdot r}}\left.  |\Psi \right\rangle \) is the quantum state of one photon under displacement \(\mathbf{r}\). Thus, \(\hat \rho_{1}\) is the mixed state of a randomly displaced photon, with the classical probability distribution \(f\left( \mathbf{r} \right)\). In this way, we convert the problem of ``imaging an arbitrary sample'' into the well-posed task of estimating the Fourier amplitudes from the quantum state of the collected photons.

For incoherent imaging, \(\left| \Psi\left( \mathbf{r} \right) \right|^{2}\) is the intensity point spread function (PSF) of DI. Its Fourier transform is the OTF: \(\beta^{\rm DI}\left( \mathbf{k} \right) = \iint_{}^{}{\left| \Psi\left( \mathbf{r} \right) \right|^{2}e^{- i\mathbf{k \cdot r}}d^{2}\mathbf{r}} = \frac{1}{4\pi^2}\widetilde{\Psi}\left( \mathbf{k} \right)*\widetilde{\Psi}\left( \mathbf{k} \right)\), where \(\widetilde{\Psi}\left( \mathbf{k} \right)\) is the Fourier transform of \(\Psi\left( \mathbf{r} \right)\), the asterisk denotes correlation. In typical microscopy, \(\widetilde{\Psi}\left( \mathbf{k} \right) = \frac{4\sqrt{\pi}}{k_c}\theta(k_c/2-\left| \mathbf{k} \right|)\) is a top-hat circular pupil function with cutoff frequency $\frac{2\pi}{\lambda}{\rm NA}=\frac{k_c}{2}$, then \(\Psi\left( \mathbf{r} \right)=\frac{J_1(k_c{r}/2)}{\sqrt{\pi}{r}}\) is an Airy disk, where \(J_1\) is the Bessel function of the first kind of order one. The QFI and FI per detected photon can be calculated approximately as (see the derivation in Supplementary Information S1) 
\begin{equation}
{\rm QFI}_{ij} \approx \frac{\beta^{\rm DI}\left( \mathbf{k}_i \right)} {2a_{0}^{2}}\delta_{ij},
\label{eq:QFI}
\end{equation}
and
\begin{equation}
{\rm FI}_{ij}^{\rm DI} \approx \frac{\left[{\beta^{\rm DI}}\left( \mathbf{k}_i \right)\right]^{2}}{2a_{0}^{2}\beta^{\rm DI}(0)} \delta_{ij}, \label{eq:FI_DI}
\end{equation}
where the parameters are represented as a single set $\{\theta_i\}=\{a_{\mathbf k}\}\bigcup\{b_{\mathbf k}\}$ and ${\mathbf k}_i$ is the wavevector associated with $\theta_i$. Both are diagonal matrices, thus \({\rm QCRB} = {\rm QFI}^{- 1}\) and \({\rm CRB}^{\rm DI} = \left({\rm FI}^{\rm DI}\right)^{- 1},\) are also diagonal. 

\subsection{Principle of Fourier domain division (FDD)}

\begin{figure*}[htbp]
  \centering
  \includegraphics[width=1\textwidth]{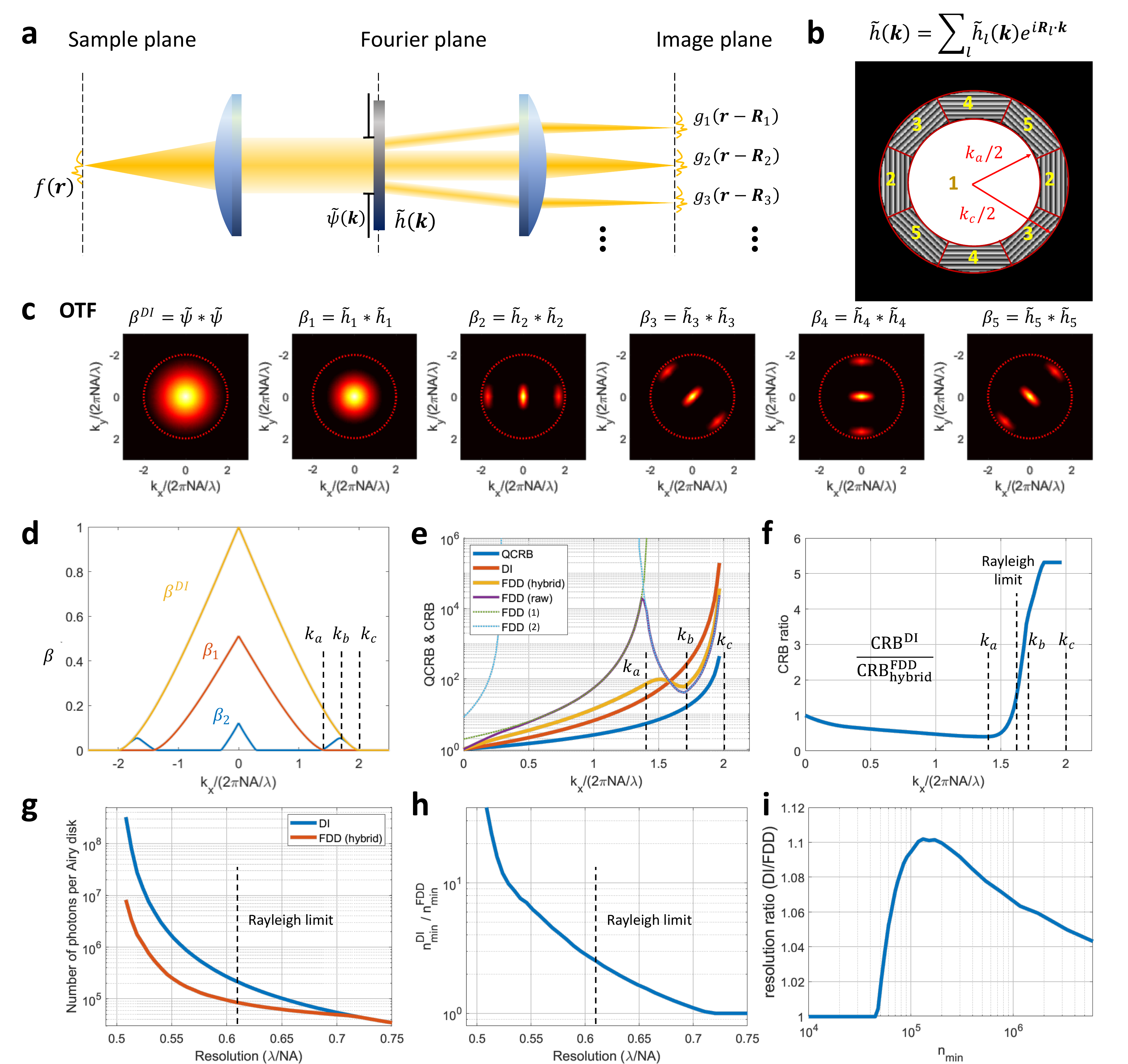}
  \caption{\textbf{Fourier Domain Division (FDD) concept. a}, Schematic of the FDD imaging system. A 4-f imaging system is modified by inserting a custom phase plate in the Fourier plane. \textbf{b,} Example phase plate dividing the pupil into five regions \({\widetilde{h}}_l\left( \mathbf{k} \right),\ \ l = 1,\ldots ,5\), each encoded with a different grating to generate five separate images \(g_l(\mathbf{r})\). \textbf{c,} Corresponding OTFs as autocorrelation of the pupil. The red dashed circle indicates the cutoff frequency \(k_c\). \textbf{d,} Line profiles of the first and the second FDD  OTFs, as well as the direct imaging OTF, along the \(k_{x}\) axis. \textbf{e,} CRBs for DI, FDD (raw), FDD (hybrid), first and the second FDD channel [FDD(1) and FDD(2)], respectively, and the QCRB; $\alpha=0.6$ is assumed. \textbf{f,} CRB ratio between DI and FDD (hybrid). For high spatial frequencies, the CRB of FDD (hybrid) is about 5 times lower than that of DI, showing significant quantum advantage. \textbf{g}, The minimum photon numbers $n_{\min}^{\rm DI}$ and $n_{\min}^{\rm FDD}$ per Airy disk area as a function of the achievable resolution. \textbf{h}, the ratio between  $n_{\min}^{\rm DI}$ and $n_{\min}^{\rm FDD}$. \textbf{i}, The ratio of achievable resolutions between DI and FDD.}
  \label{fig:1}
\end{figure*}

Eqs. (\ref{eq:QFI}), (\ref{eq:FI_DI}) show that the ratio between the FI of DI and QFI is $\beta^{\rm DI}(\mathbf{k})/\beta^{\rm DI}(0)$ for a Fourier component with frequency $\mathbf{k}$. FDD is an OTF engineering method aimed to maximize this ratio, thus enhancing FI. Fig.\ref{fig:1}(a) illustrates the principle of FDD in a standard 4-f imaging system. Light from the sample is Fourier-transformed by the first lens, truncated by a top-hat pupil function \(\widetilde{\Psi}\left( \mathbf{k} \right)\) in the Fourier plane, and then imaged onto the detector by the second lens. In the Fourier plane, we insert a custom phase plate \(\widetilde{h}\left( \mathbf{k} \right)\) that divides the support of the pupil function \(\widetilde{\Psi}\left( \mathbf{k} \right)\) into several non-overlapping regions associated with pupil functions \({\widetilde{h}}_l\left( \mathbf{k} \right)\). Each region is encoded with a distinct blazed-grating phase pattern \(e^{- i\mathbf{R}_l\mathbf{\cdot k}}\), imparting a unique transverse momentum to the light passing through it. After the second lens, these phase-encoded regions give rise to multiple displaced images \(g_l(\mathbf{r})\) in the image plane, each shifted by a displacement vector \(\mathbf{R}_l\), whose intensity distributions are measured with an array detector. In this way, information from different regions of the Fourier plane is recorded simultaneously in separate, non-overlapping images. 

Fig.\ref{fig:1}(b) shows an example phase plate where the Fourier plane is partitioned into five regions \(\widetilde{h}_l\) ($l=1,\ldots,5$), each encoded with a different grating phase to generate five corresponding images. As labeled in Fig.\ref{fig:1}(b), the radii of \(\widetilde{\Psi}\left( \mathbf{k} \right)\) and \({\widetilde{h}}_1\left( \mathbf{k} \right)\) are \(k_{c}/2\) and \(k_a/2\), respectively. The corresponding OTFs  \(\beta_l\left( \mathbf{k} \right) = \frac{1}{4\pi^2}{\widetilde{h}}_l\left( \mathbf{k} \right)*{\widetilde{h}}_l\left( \mathbf{k} \right)\) are shown in Fig.\ref{fig:1}(c).  

Fig.\ref{fig:1}(d) compares the \(k_{x}\) axis cross-sections of the OTF for DI (\(\beta^{\rm DI}\)) with that of the first and the second pupil for FDD (\(\beta_1(\mathbf{k})\) vs.~\(\beta_2( \mathbf{k})\)). Since \(\Psi\) is normalized, \(\beta^{\rm DI}(0) = \left\langle \Psi \middle| \Psi \right\rangle = 1\), and each \(\beta_l(0)\) is the area ratio between the FDD pupil function \({\widetilde{h}}_l\left( \mathbf{k} \right)\) and the DI pupil function \(\widetilde{\Psi}\left( \mathbf{k} \right)\), e.g., \(\beta_2(0) = \frac{\left( 1 - k_a^{2}/k_{c}^{2} \right)}4 \ll 1.\) On the other hand,  \(\beta^{\rm DI}\left( k_{x} \right) \approx \beta_2\left( k_{x} \right)\) for high frequencies: \(k_{x} \geq k_{b} \equiv \frac{k_a + k_{c}}{2}\). 

This behavior permits precise evaluation of the image Fourier components corresponding to these frequencies. To see this, we write the FI for the \(l^{th}\) pupil of FDD (refer to  Supplementary Information S1 for detailed calculation):
\begin{equation}
{\rm FI}_{ij}^{(l)} \approx \frac{\left[{\beta_l}\left( \mathbf{k_i} \right)\right]^2}{2a_{0}^{2}\beta_l(0)}\delta_{ij}.
\label{eq:FI_l}
\end{equation}
Because the FI scales as \(\left[{\beta_l}\left( \mathbf{k_i} \right)\right]^2/\beta_l(0)\), the reduction of \(\beta_2(0)\) boosts FI$^{(2)}$ for high-frequency components by a factor of up to \(\frac{4}{\left( 1 - k_a^{2}/k_{c}^{2} \right)}\), demonstrating a substantial quantum advantage for estimating fine spatial detail. Intuitively, detecting the regions \({\widetilde{h}}_2 - {\widetilde{h}}_5\) separately from the central region \({\widetilde{h}}_1\) allows us to measure the contribution from high-frequency Fourier components without contamination from the shot noise associated with low frequencies, therefore enhancing the precision of these measurements.

The total FI of FDD is the sum of the FIs from each pupil:
\begin{equation}
{\rm FI}_{\rm raw}^{\rm FDD} = \sum_{l = 1}^{5}{\rm FI}^{(l)}.
\label{eq:FI_raw}
\end{equation}
As seen in Fig.~\ref{fig:1}(e), \({\rm CRB}_{\rm raw}^{\rm FDD} = \left( {\rm FI}_{\rm raw}^{\rm FDD} \right)^{- 1}\) shows significant advantage with respect to \({\rm CRB}^{\rm DI}\) in the high-frequency region, with a minor loss in the low-frequency region. Unfortunately, however, it is singular for \(k = k_a\). This is because the FDD OTF at this point vanishes [Fig. \ref{fig:1}(d)]. To address this issue, we redistribute the total photon budget, allocating a fraction \(\alpha\) across the five segmented pupils, while the remaining \(1 - \alpha\) are acquired via DI. This results in the final hybrid Fisher information
\begin{equation}
{\rm FI}_{\rm hybrid}^{\rm FDD} = \alpha{\rm FI}_{\rm raw}^{\rm FDD} + (1 - \alpha){\rm FI}^{\rm DI}.
\label{eq:FI_hybrid}
\end{equation}
As shown in Fig. \ref{fig:1}(e), this hybrid strategy \({\rm CRB}_{\rm hybrid}^{\rm FDD} = \left( {\rm FI}_{\rm hybrid}^{\rm FDD} \right)^{- 1}\) suppresses the singularity, while largely maintaining the enhancement for high $\mathbf k$'s [Fig.\ref{fig:1}(f)]. 
As discussed above, the level and the frequency range of this enhancement are determined by the ratio \(k_a:k_c\). Higher ratio leads to higher enhancement but narrower range. In our experiments, we choose \(k_a=0.7k_c\), corresponding to $k_b=0.85k_c$, to cover the range of enhancement beyond the Rayleigh limit $0.61\lambda/\mathrm{NA}$, corresponding to $0.82k_c$. The photon splitting ratio $\alpha$ is chosen as $0.6$, which provides close-to-optimal resolution enhancement for our measurement regime, as discussed below.

As evidenced by the behavior of the red curve in [Fig.\ref{fig:1}(e)], the number of photons required to achieve a given precision in evaluating a particular Fourier component rapidly grows with frequency. Therefore, if our goal is to precisely evaluate a wider range of the image spectrum with a given photon budget, it is beneficial to pick a measurement that sacrifices the Fisher information for low frequencies in favor of high frequencies. This is exactly the vision of FDD.

To quantify this intuitive argument and find the achievable resolution for a given photon budget, we evaluate 
the minimum photon number $n_{\min}$ per area of an Airy disk 
such that for all $|\mathbf{k}|<k$, the ratio between the true value of the parameter $\theta_i$ and its standard deviation $\sqrt{\left< \Delta \theta_i^2 \right>}$ is above a threshold $\gamma$. As we derive in Supplementary Information S2, this minimum photon budget for DI is 
\begin{subequations}\label{eq:Nmin}
    \begin{equation}
        n_{\min}^{\rm DI}(k)\approx\frac{8\times 0.61^2\pi^2\gamma^2}{R^2}\max_{|\mathbf{k}_i|<k}\frac{\beta^{\rm DI}(0)}{\left[\beta^{\rm DI}(\mathbf{k}_i) \right]^2},
        \label{eq:N_min_DI}
    \end{equation}
where $R=\frac{ \langle\Delta f(\mathbf{r})^2\rangle}{a_0}$, a real number typically below 1, is the coefficient of variation of the sample. For hybrid FDD, it is 
    \begin{equation}
    n_{\min}^{\rm FDD}(k)\approx\frac{8\times 0.61^2\pi^2\gamma^2}{R^2}\min_{k_a, \alpha}\left[\max_{|\mathbf{k}_i|<k}\left[\sum_{l=0}^5\frac{\left[\beta^H_l(\mathbf{k}_i) \right]^2}{\beta^H_l(0)}\right]^{-1}\right],
    \label{eq:N_min_FDD}
    \end{equation}
\end{subequations}
where \(\beta^H_l=\alpha\beta_l\) for \(l=1,\ldots,5\), \(\beta^H_0=(1-\alpha)\beta^{\rm DI}\). Reversing these relations, the value of $k$ can be calculated for both methods given the photon budget $n_{\min}$. The achievable spatial resolution is then $2\pi/k$. 

Figure \ref{fig:1}(g) plots  $n_{\min}^{\rm DI}$ and $n_{\min}^{\rm FDD}$ as a function of achievable resolution $2\pi/k$, assuming $R=0.4$ and $\gamma=3$. It shows that FDD requires a much lower photon budget in the high-resolution region. The ratio $n_{\min}^{\rm DI}/n_{\min}^{\rm FDD}$ is plotted in Fig.~\ref{fig:1}(h); 2.5-fold improvement is found at the Rayleigh limit $\frac{2\pi}k=\frac{0.61\lambda}{\rm NA}$, while the improvement approaches infinity for resolution close to $\frac{2\pi}k\to\frac{2\pi}{k_c}=\frac{0.5\lambda}{\rm NA}$. Figure \ref{fig:1}(i) plots the ratio of achievable resolution between DI and FDD versus the photon budget. FDD achieves up to 1.1-fold resolution improvement.

\subsection{Image reconstruction algorithm}

We denote the total photon number as \(N\), normalized ground truth intensity distribution \(f\left( \mathbf{r} \right)\) and the recorded photon density (number of photons per area) for the $l^{th}$ pupil function as $g_l\left( \mathbf{r} \right)=\langle g_l\left( \mathbf{r} \right)\rangle+n_l(\mathbf{r})$, where $\langle g_l\left( \mathbf{r} \right)\rangle$  
is the mean value and $n_l(\mathbf{r})$ is the noise. In Fourier domain, $\tilde{g}_l\left( \mathbf{k} \right)=\langle \tilde{g}_l\left( \mathbf{k} \right)\rangle+\tilde{n}_l(\mathbf{k})$. The mean photon density is \(\langle \tilde{g}_l\left( \mathbf{k} \right)\rangle = N\beta^H_l\left( \mathbf{k} \right)\widetilde{f}\left( \mathbf{k} \right)\) , where \(\widetilde{f}\left( \mathbf{k} \right)\) is the Fourier transform of  the ground truth \(f\left( \mathbf{r} \right)\). To reconstruct the image, the six raw images produced by FDD are combined in the Fourier domain using a generalized Wiener-filter-type estimator \cite{Ref22} to yield the super-resolved reconstruction:
\begin{subequations}\label{eq:FDD_image}
\begin{equation}
{\widetilde{g}}^{\rm FDD}(\mathbf{k})=\sum_{l=0}^{5}{C_l(\mathbf{k}){\widetilde{g}}_l(\mathbf{k})}
\label{eq:FDD_image_a}   
\end{equation}
with
\begin{equation}
C_l(\mathbf{k})=\frac{{\beta^H_l(\mathbf{k})}/{\beta^H_l(0)}}{\sum_{m=0}^{5}{{\left[\beta^H_m(\mathbf{k})\right]^2}/{\beta^H_m(0)}}+\epsilon_{\rm FDD}(\mathbf{k})}, \label{eq:FDD_image_b}     
\end{equation}
where 
\begin{equation}
\epsilon_{\rm FDD}(\mathbf{k})=\frac{1}{N\left|\widetilde{f}(\mathbf{k}) \right|^{2}}     
\end{equation}
\end{subequations}
is a small regularization term, which prevents division by zero. Because this term contains the ground truth information, the estimator is applied iteratively, first initializing \(\epsilon_{\rm FDD}(\mathbf{k})=N/\left\lbrack \sum_{l}\left| {\widetilde{g}}_l(\mathbf{k}) \right| \right\rbrack^{2}\), and in subsequent iteration steps applying $\epsilon_{\rm FDD}(\mathbf{k})=\frac{N}{\left|\widetilde{g}^{\rm FDD}(\mathbf{k})\right|^{2}}$ from the output of the previous step. Finally, the reconstructed image \(g^{\rm FDD}\left( \mathbf{r} \right)\) can be obtained by inverse Fourier transform of \({\widetilde{g}}^{\rm FDD}\left( \mathbf{k} \right)\). Full details are given in Supplementary Information S3, where we derive Eqs.~(\ref{eq:FDD_image}) by minimizing the mean square error between \({\widetilde{g}}^{\rm FDD}(\mathbf{k})\) and the ground truth \(N\widetilde{f}(\mathbf{k})\). We prove that, in the limit \(\epsilon_{\rm FDD}(\mathbf{k})\rightarrow 0\), the estimator described in Eq. (\ref{eq:FDD_image}) attains the signal-to-noise ratio allowed by the CRB. 

For fair comparison with DI, we apply similar processing to the DI image, which in this case is equivalent to deconvolution by means of the conventional Wiener filter \cite{Ref22}. Because DI has only a single pupil function \(\widetilde{\Psi}\left( \mathbf{k} \right)\), Eqs.~(\ref{eq:FDD_image}) reduce to
\begin{subequations}\label{eq:DI_image}
\begin{align}
{\widetilde{g}}^{\rm DI\,dcv}(\mathbf{k})&=C^{\rm DI}(\mathbf{k}){\widetilde{g}}^{\rm DI}(\mathbf{k}), 
\label{eq:DI_image_a} \\
C^{\rm DI}(\mathbf{k})&=\frac{\beta^{\rm DI}(\mathbf{k})}{\left[\beta^{\rm DI}(\mathbf{k})\right]^2+\epsilon_{\rm DI}(\mathbf{k})}, \label{eq:DI_image_b} \\
\epsilon_{\rm DI}(\mathbf{k})&=\frac{1}{N\left|\widetilde{f}(\mathbf{k}) \right|^{2}}. 
\end{align}
\end{subequations}
This estimator features a similar circular dependency, which can be solved by the same iteration algorithm, initializing with \(\epsilon_{\rm DI}(\mathbf{k})=N/\left| {\widetilde{g}}^{\rm DI}(\mathbf{k}) \right|^{2}\).

\begin{figure}[htbp]
  \centering
  \includegraphics[width=1\columnwidth]{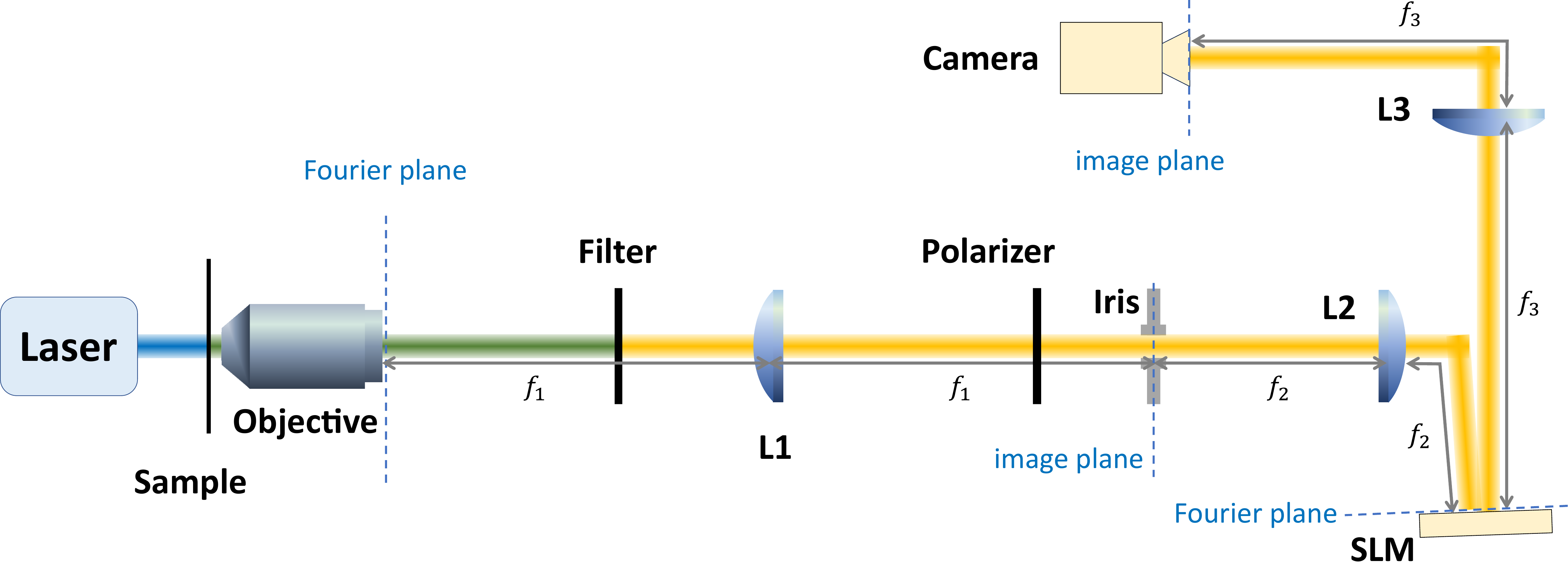}
\caption{\textbf{Experimental setup for FDD}. A CW laser (473 nm) excites the fluorescent resolution test chart. The emitted fluorescence is collected by an oil-immersion objective (100×, NA=1.4). A 4-f system consisting of lenses L1 and L2 images the Fourier plane onto the spatial light modulator (SLM), where the pupil function \(h^{(l)}\) is displayed. An iris at the intermediate image plane limits the field of view. The final image is recorded by an EMCCD camera.}
  \label{fig:3}
\end{figure}

\subsection{Experimental validation of FDD}

The experimental setup is shown in Fig.~\ref{fig:3}. A continuous-wave (CW) laser at 473 nm (Toptica iBeam smart-473) illuminates a fluorescent resolution test chart (TC-RT03fl, Technologie Manufaktur GmbH \& Co. KG), which consists of parallel line quintuplets with spatial frequencies labelled in lines/mm. The emitted fluorescence is collected by an oil-immersion objective lens (UPlanSApo 100×, NA~1.4, Olympus). The back aperture of the objective corresponds to the Fourier plane of the imaging system. A bandpass filter (FBH540-10, Thorlabs) is used to isolate the fluorescence from the laser. A 4-f optical system, formed by lenses L1 (\(f_1=500\)~mm) and L2 (\(f_2=300\)~mm), images this Fourier plane onto a spatial light modulator (SLM, Meadowlark E19x12-500-1200-HDM8), where the pupil function \(\widetilde h(\mathbf{k})\) is displayed. An iris is placed at the intermediate image plane to control the field of view, enabling discretization of $\mathbf k$ in Eq.~\eqref{eq:sample}. A polarizer is used to select the vertically polarized component, to match the modulation polarization of the SLM. The final image is relayed by lens L3 (\(f_3=500\)~ mm) and recorded using an EMCCD camera (iXon 885K, Andor). For the DI image, we display a blank hologram on the SLM, so the SLM serves as a mirror.

\begin{figure*}[htbp]
  \centering
  \includegraphics[width=0.8\textwidth]{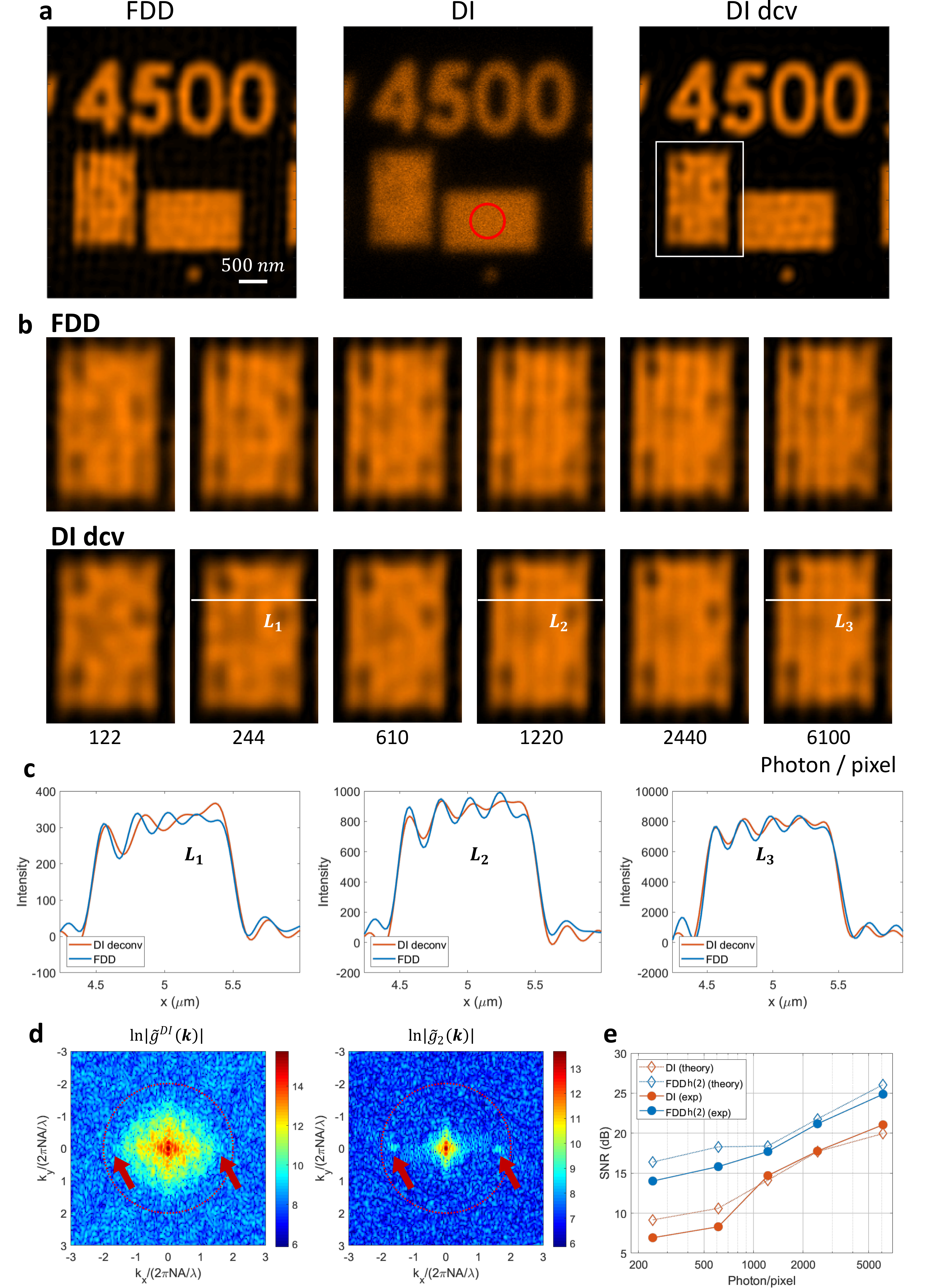}
\caption{\textbf{Experimental validation of FDD. a}, Comparison of FDD, DI, and DI deconvolution (DI dcv) for a fluorescent resolution chart (4500 lines/mm). Within the red dashed circle, the average photon number is 244 photons/pixel. DI is dominated by noise, while FDD resolves five lines. \textbf{b}, Zoomed-in regions (white box in \textbf{a}) at different photon budgets; FDD resolves the lines at 244 photons/pixel, whereas DI dcv requires $\ge1220$ photons/pixel. \textbf{c}, Line profiles along L$_1$--L$_3$ labelled in \textbf{b}. \textbf{d}, Fourier spectra of DI and FDD raw data of the second pupil \(h_2\); high-frequency components (red arrows) are visible in FDD but buried in noise in DI. \textbf{e}, SNR of the 4500 lines/mm component ($|\mathbf{k}|=0.87 k_c$) versus photon budget; FDD shows 4.97 dB improvement on average.}
  \label{fig:2}
\end{figure*}

With NA=1.4 and the  fluorescent wavelength \(\lambda=540\) nm, the cutoff frequency of our system \(k_c=\frac{4{\pi}NA}{\lambda}\) corresponds to 5184 lines/mm. The Rayleigh criterion (\(\frac{0.61\lambda}{NA}\)=235 nm) corresponds to 4255 lines/mm. The sample frequency of 4500 lines/mm is beyond the Rayleigh criterion but within the cutoff frequency. A field of view of $276\times264$ pixels is selected, with the pixel size corresponding to $16.7\ \rm nm$. Fig.~\ref{fig:2}(a) compares FDD, DI, and deconvolved DI (DI dcv) images acquired with 244 photons per pixel ($\approx 1.5\times10^5$ photons per Airy disk, measured within the red dashed circle), where the FDD and DI dcv images are obtained via Eqs.~(\ref{eq:FDD_image}) and (\ref{eq:DI_image}), respectively, with three iterations. The DI and DI dcv images appear dominated by noise, the lines cannot be distinguished, and deconvolution amplifies the high frequency noise. In contrast, FDD under the same photon budget clearly resolves the five parallel lines. 

Notably, the horizontal line quintuplet is not resolved by any of the methods. This is because the vector nature of the field causes the PSF to become slightly wider in the dimension parallel to the polarization direction under a high-NA objective lens \cite{Ref27}, which is vertical in our case.

A systematic comparison of FDD and DI dcv under different photon budgets is shown in Figs. \ref{fig:2}(b,c), where the region in the white box in Fig. \ref{fig:2}(a) is selected for comparison. For FDD, 244 photons/pixel are sufficient to resolve the five-line structure, whereas DI dcv requires at least 1220 photons/pixel to achieve comparable resolution. At high photon budget, both methods clearly recover the five lines with similar quality.

Figure \ref{fig:2}(d) illustrates the advantage of FDD in the Fourier domain. The red arrows indicate high-frequency components corresponding to the 4500 lines/mm structure. In the FDD data from the second pupil ($h_2$) [Fig.~\ref{fig:2}(d), right], these components are clearly visible, whereas in the DI case [Fig.~\ref{fig:2}(d), left] they are buried in noise due to the higher shot noise level (brighter background). Finally, the signal-to-noise ratio (SNR) of the Fourier component at 4500 lines/mm is plotted in Fig.~\ref{fig:2}(e). Noise of a raw image was quantified in two ways: (i) theoretically, using the shot-noise model \(\langle\left|\tilde{n}_l(\mathbf{k})\right|^2\rangle=N_l\) [Eq. (\ref{eq:noise}) in S3], where $N_l$ is the total number of photons in the $l$th pupil; (ii) experimentally, by averaging the intensity outside the cutoff frequency in the Fourier domain, where no signal is present [Fig.~\ref{fig:2}(d)], according to $\langle\left|\tilde{n}_l(\mathbf{k})\right|^2\rangle=\frac{\int_{\mathbf{k}>k_c}|\tilde{g_l}(\mathbf{k})|^2d^2\mathbf{k}}{\int_{\mathbf{k}>k_c}d^2\mathbf{k}}$. Compared with DI, FDD should theoretically achieve an SNR improvement of 6.63 dB on average for $|\mathbf{k}|\sim k_b$. In  experiment, the SNR  improves by 4.97 dB. The discrepancy is due to electronic noise.

\section{Discussion}

FDD shows significant quantum advantage in terms of photon budget, making it ideal for imaging scenarios under limited photon numbers. Instances include photo-toxicity in bio-imaging \cite{Ref23}, observation of faint celestial bodies in astronomy, and high speed photography with extremely short exposure times. In addition, FDD is a passive method, which makes it broadly applicable in contexts where active interaction with the sample is impossible, including astronomy, satellite imaging, and surveillance. 

The novelty of our method stems from its foundation in quantum estimation theory. By analytically calculating QFI and FI, we convert the quantum optimization task to a classical OTF engineering problem --- more specifically, optimizing the ratio $\beta(\mathbf{k})/\beta(0)$.  This enables relatively straightforward extension of our approach to other imaging modalities, beyond incoherent linear imaging under wide-field illumination studied here.  
\begin{itemize}
    \item For nonlinear imaging methods, such as two-photon fluorescence, the OTF deviates from a triangular shape, exhibiting a sharp peak around zero and quickly falling off for higher frequency components \cite{Ref29}. Hence we expect that FDD, which reduces the denominator $\beta(0)$, will be even more beneficial than in the linear case.
    \item Prior knowledge of the sample, e.g.,~limited size, helps to retrieve the frequency components beyond the cutoff. However, the accuracy of the retrieval depends on the SNR \cite{Ref28}. Since our method improves the SNR significantly for high frequency components close to the cutoff, it is interesting to explore the QFI, the FI and the optimal detection in this scenario. 
    \item If a structured pattern of illumination is an option, this gives us another degree of freedom, which affects the quantum state of light entering the objective lens, and hence its QFI itself. By optimizing the illumination structure in concert with the OTF, further resolution enhancements can be expected.
\end{itemize}


\section{Acknowledgments}

The project is funded by EPSRC Standard Grant EP/Y020596/1 and  EPSRC Impact Acceleration Account
Award EP/X525777/1. A.~Z.~acknowledges a UK Research and Innovation Guarantee Postdoctoral Fellowship under the UK government’s Horizon Europe funding Guarantee (Q-LIMAGE, EP/Y029127/1).

\bibliographystyle{unsrt}  
\bibliography{reference}  

\newpage
\section*{Supplimentary Information}
\subsection{S1. QFI and FI analysis.}
In this subsection, we derive approximate analytical solutions for the QFI, the FI of DI, and the FI of FDD. These analytical results are then validated by comparing  with numerical solutions obtained without approximation.  

\subsubsection{Quantum Fisher information}
The quantum Fisher information per photon can be calculated as \cite{Ref26}
\begin{equation}
{\rm QFI}_{ij} = {\rm Re}\left\lbrack {\rm tr}\left( \hat\rho_{1}\hat L_i\hat L_j \right) \right\rbrack,
\label{eq:QFI0}
\end{equation}
where
\begin{equation}
\frac{{\partial\hat\rho}_{1}}{\partial \theta_i} =\frac{1}{2} \left(\hat \rho_{1} \hat L_i+ \hat L_i\hat\rho_{1} \right).
\label{eq:SLD0}
\end{equation}
defines the symmetric logarithmic derivative (SLD) \(\hat L_i\). Equations (\ref{eq:sample}) and  (\ref{eq:density_matrix}) cannot be solved analytically, so we introduce the following approximation 
\begin{equation}
\hat\rho_{1}^{(\mathrm{appr})}= \iint{a_0e^{- i\mathbf{\hat K \cdot r}}\left. \ |\Psi \right\rangle\left\langle \Psi| \right.\ e^{ i\mathbf{\hat K \cdot r}}d^{2}\mathbf{r}}.
\label{eq:rho_appr}
\end{equation}
\emph{only for the right-hand side} of Eq.
~(\ref{eq:SLD0}). In other words, the intensity is assumed spatially independent. 

Using the original density matrix for the left-hand side, we find  
\begin{subequations}\label{eq:drho_da}
\begin{align}
\frac{{\partial\hat\rho}_{1}}{\partial a_\mathbf{k}} &=  \iint_{}^{}{\cos\left( \mathbf{k \cdot r} \right)e^{- i\mathbf{\hat K \cdot r}}\left. \ |\Psi \right\rangle\left\langle \Psi| \right.\ e^{ i\mathbf{\hat K \cdot r}}d^{2}\mathbf{r}}\\
\frac{{\partial\hat\rho}_{1}}{\partial b_\mathbf{k}} &=  \iint_{}^{}{\sin\left( \mathbf{k \cdot r} \right)e^{- i\mathbf{\hat K \cdot r}}\left. \ |\Psi \right\rangle\left\langle \Psi| \right.\ e^{ i\mathbf{\hat K \cdot r}}d^{2}\mathbf{r}}.
\end{align}
\end{subequations}
Then we obtain the approximation of SLD as
\begin{equation}
\hat L_i^{(\mathrm{appr})}=\frac{k_c^2}{16\pi a_0}\frac{{\partial\hat\rho}_{1}}{\partial \theta_i} .
\label{eq:SLD}
\end{equation}
The derivation of this result is tedious. However, it can be verified as follows. From Eqs.~\eqref{eq:rho_appr} and \eqref{eq:SLD} we have 
\begin{equation}
\begin{split}
& \hat\rho_{1}^{(\mathrm{appr})}\hat  L_{a_\mathbf{k}}^{(\mathrm{appr})}= \frac{k_c^2}{16\pi a_0}\iint{a_0e^{- i\mathbf{\hat K \cdot r}} |\Psi\rangle\langle \Psi|  e^{ i\mathbf{\hat K \cdot r}}d^{2}\mathbf{r}}\\
&\times\iint_{}^{}{\cos\left( \mathbf{k\cdot r'} \right)e^{- i\mathbf{\hat K \cdot r'}}|\Psi\rangle\langle \Psi|e^{ i\mathbf{\hat K \cdot r'}}d^{2}\mathbf{r'}}.\\
&=\frac{k_c^2}{16\pi }\iint{d^2\mathbf{r'}\cos( \mathbf{k \cdot r'})}\\
&\times\left[\iint{d^2\mathbf{r}}\left\langle \Psi|e^{i\mathbf{\hat K \cdot r}}e^{- i\mathbf{\hat K \cdot r'}}|\Psi \right\rangle e^{- i\mathbf{\hat K \cdot r}}|\Psi\rangle\right] \langle \Psi| e^{ i\mathbf{\hat K \cdot r'}}
\end{split}
\end{equation}

The term in angular brackets in the last line is a self-convolution of the APSF: 
\begin{equation}
\begin{split}
 &\left\langle \Psi|e^{i\mathbf{\hat K \cdot r}}e^{- i\mathbf{\hat K \cdot r'}}|\Psi \right\rangle=\iint{\langle \Psi|e^{i\mathbf{\hat K \cdot r}}|\mathbf{r_0}\rangle\langle\mathbf{r_0}|e^{- i\mathbf{\hat K \cdot r'}}|\Psi\rangle}d^2\mathbf{r_0}\\
 &=\iint{\Psi(\mathbf{r-r_0})\Psi(\mathbf{r'-r_0})}d^2\mathbf{r_0}=\frac{4\sqrt{\pi}}{k_c}\Psi(\mathbf{r-r'}).
 \end{split}
\end{equation}
In the last equality, we used the fact that 
\(\Psi\left( \mathbf{r} \right)=\frac{J_1(k_c\mathbf{r}/2)}{\sqrt{\pi}\mathbf{r}}\) is a normalized Airy disk, so its self convolution is also an Airy disk, i.e., \(\iint{\Psi\left( \mathbf{r} \right)\Psi\left( \mathbf{r'-r} \right)}d^2\mathbf{r}=\frac{4\sqrt{\pi}}{k_c}\Psi\left( \mathbf{r'} \right)\). This is because the Fourier transform of the Airy disk is a top-hat function, and its self-product is also a top-hat function.

\begin{widetext}
Hence
\begin{equation}
\begin{split}
\hat\rho_{1}^{(\mathrm{appr})}\hat L_{a_\mathbf{k}}^{(\mathrm{appr})}&=\frac{k_c^2}{16\pi }\iint{d^2\mathbf{r'}\cos( \mathbf{k \cdot r'})\left[\iint{d^2\mathbf{r}}\frac{4\sqrt{\pi}}{k_c}\Psi(\mathbf{r-r'})e^{- i\mathbf{\hat K \cdot r}}\left. \ |\Psi \right\rangle\right] \langle \Psi| e^{ i\mathbf{\hat K \cdot r'}}}\\
&=\frac{k_c^2}{16\pi }\iint{d^2\mathbf{r'}\cos( \mathbf{k \cdot r'})\left[\iint{d^2\mathbf{r}}\frac{4\sqrt{\pi}}{k_c}\Psi(\mathbf{r-r'})\iint{d^2\mathbf{r_0}}|\mathbf{r_0}\rangle\langle\mathbf{r_0}|e^{- i\mathbf{\hat K \cdot r}}\left. \ |\Psi \right\rangle\right] \langle \Psi| e^{ i\mathbf{\hat K \cdot r'}}}\\
&=\frac{k_c^2}{16\pi }\iint{d^2\mathbf{r'}\cos( \mathbf{k \cdot r'})\left[\frac{4\sqrt{\pi}}{k_c}\iint{d^2\mathbf{r_0}}|\mathbf{r_0}\rangle\iint{d^2\mathbf{r}}\Psi(\mathbf{r-r'})\Psi(\mathbf{r-r_0})\right] \langle \Psi| e^{ i\mathbf{\hat K \cdot r'}}}\\
&=\iint{d^2\mathbf{r'}\cos( \mathbf{k \cdot r'})\left[\iint{d^2\mathbf{r_0}}|\mathbf{r_0}\rangle\Psi(\mathbf{r'-r_0})\right] \langle \Psi| e^{ i\mathbf{\hat K \cdot r'}}}\\
&=\iint{d^2\mathbf{r'}\cos( \mathbf{k \cdot r'}) e^{- i\mathbf{\hat K \cdot r'}}\left. \ |\Psi \right\rangle\langle \Psi| e^{ i\mathbf{\hat K \cdot r'}}}=\frac{{\partial\hat\rho}_{1}}{\partial a_\mathbf{k}},
\end{split}
\end{equation}
where we again used the aforementioned identity for the self-convolution of the Airy disk.
\end{widetext}

Similarly, we have \(\hat L_{a_\mathbf{k}}^{(\mathrm{appr})}\hat\rho_{1}^{(\mathrm{appr})}=\frac{{\partial\hat\rho}_{1}}{\partial a_\mathbf{k}} \). Therefore, we have verified that the SLD \(\hat L_{a_\mathbf{k}}^{(\mathrm{appr})}\) in Eq. (\ref{eq:SLD}) is a solution of Eq. (\ref{eq:SLD0}). The same is true for the sin mode \(\hat L_{b_\mathbf{k}}^{(\mathrm{appr})}\). Then, by replacing \(\hat L_i\) with \(\hat  L_i^{(\mathrm{appr})}\) and  \(\hat\rho_1\) with \(\hat\rho_{1}^{(\mathrm{appr})}\) in Eq. (\ref{eq:QFI0}) , we obtain
\begin{equation}
\begin{split}
&{\rm QFI}_{a_\mathbf{k},a_\mathbf{k}} \approx {\rm Re}\left\lbrack {\rm tr}\left( \hat\rho_{1}^{(\mathrm{appr})}\hat L_{a_\mathbf{k}}^{(\mathrm{appr})}\hat L_{a_\mathbf{k}}^{(\mathrm{appr})} \right) \right\rbrack\\
&=\frac{k_c^2}{16\pi a_0}{\rm Re}\left\lbrack {\rm tr}\left( \frac{{\partial\hat\rho}_{1}}{\partial a_\mathbf{k}}\frac{{\partial\hat\rho}_{1}}{\partial a_\mathbf{k}} \right) \right\rbrack\\
&=\frac{1}{a_0}\iint{d^2\mathbf{r}}\iint{d^2\mathbf{r'}}\cos\left( \mathbf{k \cdot r} \right)\cos\left( \mathbf{k \cdot r'} \right)\left[\Psi(\mathbf{r-r'})\right]^2\\
&=\frac{\beta^{\rm DI}\left( \mathbf{k} \right)}{a_0}\iint{d^2\mathbf{r}}\cos^2\left( \mathbf{k \cdot r} \right)\\
&=\frac{\beta^{\rm DI}\left( \mathbf{k} \right)}{2a_{0}^{2}}.
\label{eq:QFI1}
\end{split}
\end{equation}
In the last two steps, we have used \(\iint{d^2\mathbf{r'}}\cos\left( \mathbf{k \cdot r'} \right)\left[\Psi(\mathbf{r-r'})\right]^2=\beta^{\rm DI}(\mathbf{k})\cos\left( \mathbf{k \cdot r} \right)\) and \(\iint{d^2\mathbf{r}}\cos^2\left( \mathbf{k \cdot r} \right)=\frac{A}{2}\), where \(A\) is the area of the sample. The normalization condition \(\iint_{}^{}{f(\mathbf{r})\ d^{2}\mathbf{r}} = 1\) leads to \(A=1/a_0\). Similar procedures show that the diagonal elements for sin modes \({\rm QFI}_{b_\mathbf{k},b_\mathbf{k}}\approx{\rm QFI}_{a_\mathbf{k},a_\mathbf{k}}\), because  \(\iint{d^2\mathbf{r}}\sin^2\left( \mathbf{k \cdot r} \right)=\frac{A}{2}\) as well. There are two types of off-diagonal elements in QFI matrix: (1) interaction between cos and sin modes, i.e., \({\rm QFI}_{a_\mathbf{k},b_\mathbf{k'}}\) and \({\rm QFI}_{b_\mathbf{k'},a_\mathbf{k}}\), which are proportional to \(\iint{d^2\mathbf{r}}\cos\left( \mathbf{k \cdot r} \right)\sin\left( \mathbf{k' \cdot r} \right)=0\); (2) interaction between two cos or sin modes with different frequencies, i.e., \({\rm QFI}_{a_\mathbf{k},a_\mathbf{k'}}\) and \({\rm QFI}_{b_\mathbf{k},b_\mathbf{k'}}\) with \(\mathbf{k}\neq
\mathbf{k'}\), which are proportional to \(\iint{d^2\mathbf{r}}\cos\left( \mathbf{k \cdot r} \right)\cos\left( \mathbf{k' \cdot r} \right)=\iint{d^2\mathbf{r}}\sin\left( \mathbf{k \cdot r} \right)\sin\left( \mathbf{k' \cdot r} \right)=0\). As a result, we can express the approximate solution of QFI matrix as Eq. (\ref{eq:QFI}).

\subsubsection{Fisher information for DI and FDD}
With the total photon number \(N\) and normalized ground truth \(f\left( \mathbf{r} \right)\), the recorded photon density in the DI image is $g^{\rm DI}\left( \mathbf{r} \right)=\langle g^{\rm DI}\left( \mathbf{r} \right)\rangle+n^{\rm DI}(\mathbf{r})$, where $\langle g^{\rm DI}\left( \mathbf{r} \right)\rangle$ is the mean value, $n^{\rm DI}(\mathbf{r})$ is the noise. In the Fourier domain, $\tilde{g}^{\rm DI}\left( \mathbf{k} \right)=\langle \tilde{g}^{\rm DI}\left( \mathbf{k} \right)\rangle+\tilde{n}^{\rm DI}(\mathbf{k})$, with \(\langle {\widetilde{g}}^{\rm DI}\left( \mathbf{k} \right)\rangle  = N\beta^{\rm DI}\left( \mathbf{k} \right)\widetilde{f}\left( \mathbf{k} \right) \). Thus, in spatial domain, it is 
\begin{align}
\langle g^{\rm DI}\left( \mathbf{r} \right)\rangle & = N\beta^{\rm DI}(0) a_{0}  \\
&+N\sum_{\substack{|\mathbf{k}|\leq k_c\\ k_x,k_y > 0}}^{}{\beta^{\rm DI}(\mathbf{k})\left[a_{\mathbf{k}}\cos\left( \mathbf{k \cdot r} \right) + b_{\mathbf{k}}\sin\left( \mathbf{k \cdot r} \right)\right]}
\label{eq:DI_raw}
\end{align} 
Then, the mean photon number on the $i^{th}$ pixel at $\mathbf{r}_i$ is $\langle g^{\rm DI}\left( \mathbf{r}_i \right)\rangle\Delta x^2$,  where $\Delta x^2$ is the area of the pixel. Here we assume that the pixel size of the detector is much smaller than the size of the PSF. Assuming the detection is limited by the shot noise, which follows the Poisson distribution, the classical FI per photon of DI can be calculated as \cite{Ref7}
\begin{equation}
{\rm FI}_{ij}^{\rm DI} =\frac{1}{N}\iint{\frac{1}{\langle g^{\rm DI}\left( \mathbf{r} \right)\rangle}\frac{\partial\langle g^{\rm DI}\left( \mathbf{r} \right)\rangle}{\partial \theta_i}\frac{\partial\langle g^{\rm DI}\left( \mathbf{r} \right)\rangle}{\partial \theta_j}d^{2}\mathbf{r}},
\label{eq:FI_DI0}
\end{equation}
In this integral, \(\frac{\partial\langle g^{\rm DI}\left( \mathbf{r} \right)\rangle}{\partial \theta_i}\) can be evaluated exactly as 
\begin{subequations}
    \begin{align}
        \frac{\partial\langle g^{\rm DI}\left( \mathbf{r} \right)\rangle}{\partial a_\mathbf{k}}=N\beta^{\rm DI}(\mathbf{k})\cos\left( \mathbf{k \cdot r} \right) ,\\
        \frac{\partial\langle g^{\rm DI}\left( \mathbf{r} \right)\rangle}{\partial b_\mathbf{k}}=N\beta^{\rm DI}(\mathbf{k})\sin\left( \mathbf{k \cdot r} \right),
    \end{align}
\end{subequations}
while $\frac{1}{\langle g^{\rm DI}\rangle}$ can be evaluated approximately as \(\frac{1}{N\beta^{\rm DI}(0)a_0}\). This approximation is a classical version of the approximation applied to the calculation of QFI above. Then the FI of DI can be calculated analytically as
\begin{equation}
\begin{split}
    {\rm FI}_{a_\mathbf{k},a_\mathbf{k}}^{\rm DI}& \approx\frac{1}{N}\iint{\frac{1}{N\beta^{\rm DI}(0)a_0}\left[N\beta^{\rm DI}(\mathbf{k})\cos\left( \mathbf{k \cdot r} \right)\right]^2d^{2}\mathbf{r}},\\
    &=\frac{\left[\beta^{\rm DI}(\mathbf{k})\right]^2}{\beta^{\rm DI}(0)a_0}\iint{\cos^2\left( \mathbf{k \cdot r} \right)d^2\mathbf{r}}\\
    &=\frac{\left[\beta^{\rm DI}(\mathbf{k})\right]^2}{2\beta^{\rm DI}(0)a_0^2}
\label{eq:FI_DI1}
\end{split}
\end{equation}
Again, in the last step, we have used \(\iint{\cos^2\left( \mathbf{k \cdot r} \right)d^2\mathbf{r}}=\frac{A}{2}=\frac{1}{2a_0}\).  Since \(\beta^{\rm DI}(0)=1\), \({\rm FI}_{a_\mathbf{k},a_\mathbf{k}}^{\rm DI}= \frac{\left[\beta^{\rm DI}(\mathbf{k})\right]^2}{2a_0^2}\). Similarly, the diagonal element for the sin mode is \({\rm FI}_{b_\mathbf{k},b_\mathbf{k}}^{\rm DI}= \frac{\left[\beta^{\rm DI}(\mathbf{k})\right]^2}{2a_0^2}\) as well. All the off-diagonal elements are zero. Hence, we can express FI matrix as Eq. (\ref{eq:FI_DI}).

\emph{The FI of FDD} for the \(l^{th}\) pupil function \(h_l\) can be calculated via the same procedure from Eq. (\ref{eq:DI_raw}) to Eq. (\ref{eq:FI_DI1}), by replacing \(\beta^{\rm DI}\) with \(\beta_l\), resulting in Eq.~(\ref{eq:FI_l}). 

\subsubsection{Numerical calculation of quantum and classical FI}
The analytical results of QFI and FI have been obtained with an approximation. In order to verify the validity of the approximations employed, we compare the analytical results with numerical solutions obtained without approximation. The numerical FI for the DI case can be computed using Eqs.~(\ref{eq:DI_raw})--(\ref{eq:FI_DI0}) above. For the FDD case, the same equations apply with \(\beta^{\rm DI}\) replaced by \(\beta_l\). The numerical QFI can be evaluated as follows. Given the spectral decomposition of the single-photon density matrix
\begin{equation}
\hat\rho_1 = \sum_{l \in S} \xi_l \, |\psi_l\rangle \langle \psi_l|,
\end{equation}
where \( S = \{ l \, | \, \xi_l \neq 0 \} \) is the support, an element of the QFI can be calculated as \cite{Ref26}
\begin{align}
QFI_{ij}
&= \sum_{l \in S} 
   \frac{(\partial_{i} \xi_l)(\partial_{j} \xi_l)}{\xi_l}
   + \sum_{l \in S} 
   4 \xi_l \, \mathrm{Re}\!\left(
       \langle \partial_{i} \psi_l | 
       \partial_{j} \psi_l \rangle
     \right) \nonumber \\
&\quad - \sum_{l,m \in S} 
   \frac{8 \xi_l \xi_m}{\xi_l + \xi_m} \,
   \mathrm{Re}\!\left(
     \langle \partial_{i} \psi_l | \psi_m \rangle
     \langle \psi_m | \partial_{j} \psi_l \rangle
   \right).
   \label{eq:QFI_numerical}
\end{align}

We compare the analytical and numerical QFI calculations for a one-dimensional sample, such that \(f(x) = a_{0} + \sum_{l}^{}{a_l\cos\left( k_lx \right) + b_l\sin\left( k_lx \right)}\), where \(a_l\) and \(b_l\) (\(1\le l \le 48\)) are random numbers sampled from a Gaussian distribution, with \(a_0\) is chosen to make \(f(x)\) positive with \(\min[f(x)]=0.1\,\max[f(x)]\).  \\

\begin{figure}[htbp]
 \centering
\includegraphics[width=\columnwidth]{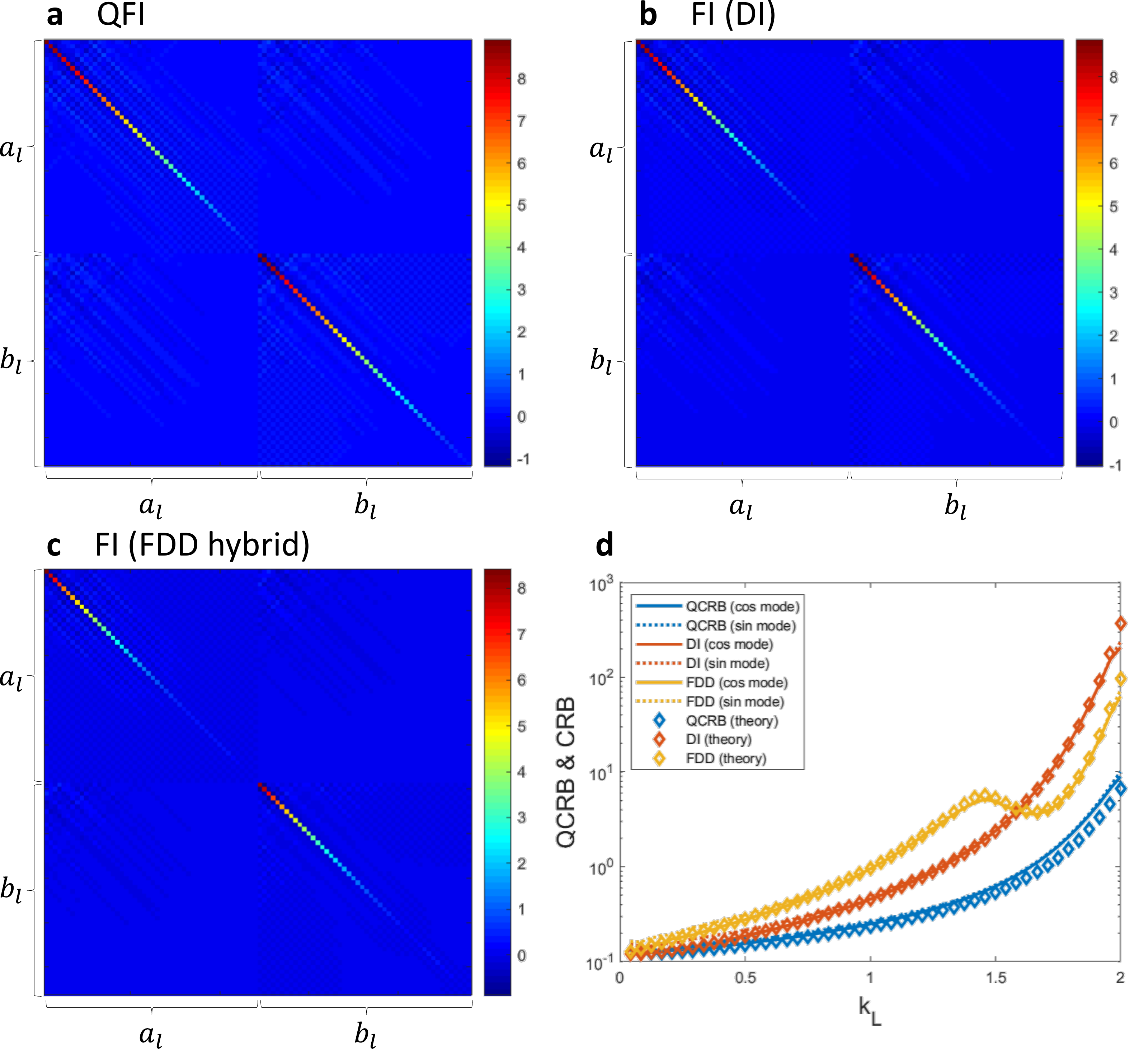}
\caption{\textbf{Comparison between approximate analytical and numerical results of QFI and FI. a-c}, numerically computed QFI, FI (DI), and FI (FDD hybrid). The matrices are nearly diagonal, consistent with the approximate analytical solutions, which are strictly diagonal. \textbf{d}, Comparison of the numerical QCRB and CRBs for cosine (solid curve) and sine (dashed curve) modes with the analytical results (diamonds). The analytical solutions yield identical results for the two modes. The good agreement verifies the validity of the approximations used in the analytical analysis.}
 \label{fig:4}
 \end{figure}

Fig. \ref{fig:4} presents a comparison between the analytical and numerical results. Figs. \ref{fig:4}(a–c) display the numerically computed QFI, FI (DI), and FI (FDD hybrid) matrices, respectively. It shows that these matrices are nearly diagonal, consistent with the analytical results in which the corresponding matrices are exactly diagonal. It also shows that the behavior of cos  ($a_l$) and sin ($b_l$) modes are almost the same, consistent with the analytical prediction where they are exactly the same. This demonstrates that the approximations used in the analytical derivations accurately capture the essential behavior of the system. In Fig. \ref{fig:4}(d), the numerical diagonal elements of QCRB and CRB matrices  for cos and sin modes are plotted as solid and dashed curves, respectively, while the analytical solutions, which are identical for the two modes, are plotted as diamonds. The agreement between the numerical and analytical results confirms the validity of the approximations employed.

\subsection{S2. Minimum photon number for achievable resolution}
To derive Eq. (\ref{eq:Nmin}), we assume that the variance of the parameter saturates the CRB, i.e.,

\begin{equation}
    \left< \Delta \theta_i^2 \right>=\frac{CRB_{i,i}}{N}
\end{equation}
For DI (\ref{eq:FI_DI}), the per-photon ${\rm CRB}^{\rm DI}_{ii}=\frac{2\beta^{\rm DI}(0)a_0^2}{\left[\beta^{\rm DI}(\mathbf{k_i})\right]^2}$, so achieving the signal-to-noise threshold $\gamma=\min_i\theta_i/\sqrt{\Delta\theta_i^2}$  requires 
\begin{equation}
    \frac{\theta_i^2}{\left< \Delta \theta_i^2 \right>}=\frac{\theta_i^2\left[\beta^{\rm DI}(\mathbf{k_i})\right]^2}{2N\beta^{\rm DI}(0)a_0^2}\ge \gamma^2,
\end{equation}
Thus, the minimum total photon number is
    \begin{equation}
        N_{\min}^{\rm DI}=\max_{|\mathbf{k}_i|<k}\frac{2a_0^2}{\theta_i^2}\frac{\beta^{\rm DI}(0)}{\left[\beta^{\rm DI}(\mathbf{k}_i) \right]^2}\gamma^2.
        \label{eq:N_min2}
    \end{equation}
This number is inconvenient for two reasons: it scales with the sample area and we have no information about individual Fourier coefficients $\theta_i$. To address the latter point, we replace in Eq.~\eqref{eq:N_min2} $\theta_i^2$ by $\bar\sigma^2\equiv\langle\theta_i^2\rangle_{\mathrm{k}}$ (i.e.~assuming a white spectrum of $f(\mathbf{r}$). The parameter $\bar\sigma$ can be related to the properties of the sample as follows.

The variance of the sample function can be calculated as 
\begin{equation}
\begin{split}
    \langle\Delta f(\mathbf{r})^2\rangle&=\langle f(\mathbf{r})^2\rangle-\langle f(\mathbf{r})\rangle^2\\
    &=\sum_{\mathbf{k}}\langle a_{\mathbf{k}}^2\rangle\rm{cos}^2(\mathbf{k \cdot r})+\langle b_{\mathbf{k}}^2\rangle\rm{sin}^2(\mathbf{k \cdot r})\\
    &=N_k\bar\sigma^2,
\end{split}
\end{equation}
where $N_k$ is the number of Fourier modes. Finite sample size leads to discrete Fourier domain sampling with a step size of $\Delta k=2\pi/\sqrt{A}$, thus within the cutoff frequency, the number of Fourier modes is $N_k=\frac{\pi k_c^2}{\Delta k^2}=\frac{k_c^2A}{4\pi}$. Let's define the coefficient of variation of the sample as
\begin{equation}
    R=\frac{ \sqrt{\langle\Delta f(\mathbf{r})^2\rangle}}{a_0}=\frac{k_c\bar\sigma}{2\sqrt{\pi}}A^{\frac{3}{2}}.
    \label{eq:roughness}
\end{equation}
Equation \eqref{eq:N_min2} then becomes 
    \begin{equation}
        N_{\min}^{\rm DI}\approx\frac{k_c^2}{2\pi R^2}A\max_{|\mathbf{k}_i|<k}\frac{\beta^{\rm DI}(0)}{\left[\beta^{\rm DI}(\mathbf{k}_i) \right]^2}\gamma^2,
        \label{eq:N_min2}
    \end{equation}
where we used $a_0=1/A$. The right-hand side is now explicitly proportional to the sample area because  $R$ is independent of it. 

Finally, we normalize the photon number to the Airy disk area  $\pi\left(\frac{0.61\lambda}{\rm NA}\right)^2$, obtaining
    \begin{equation}
        n_{\min}^{\rm DI}\approx\max_{|\mathbf{k}_i|<k}\frac{8\times 0.61^2\pi^2}{R^2}\frac{\beta^{\rm DI}(0)}{\left[\beta^{\rm DI}(\mathbf{k}_i) \right]^2}\gamma^2,
        \label{eq:N_min3}
    \end{equation}
which is Eq. (\ref{eq:N_min_DI}) in the main text. For hybrid FDD, similar analysis leads to 
    \begin{equation}
    n_{\min}^{\rm FDD}(\alpha,k_a)\approx\max_{|\mathbf{k}_i|<k}\frac{8\times 0.61^2\pi^2}{R^2}\left[\sum_{l=0}^5\frac{\left[\beta^H_l(\mathbf{k}_i) \right]^2}{\beta^H_l(0)}\right]^{-1}\gamma^2,
    \end{equation}
which is a function of $\alpha$ and $k_a$. The overall minimum photon number should be optimized over $\alpha$ and $k_a$, thus leads to Eq. (\ref{eq:N_min_FDD}) in the main text.

\subsection{S3. Image reconstruction algorithm for FDD and DI dcv.}

\subsubsection{Derivation of image reconstruction algorithm}
For FDD, our task is to find the coefficients \(C_l(\mathbf{k})\) for linear image reconstruction
$\widetilde{g}^{\rm FDD}(\mathbf{k}) = \sum_{l=0}^{5} C_l(\mathbf{k}) \, \tilde{g}_l(\mathbf{k})$ (Eq. (\ref{eq:FDD_image_a})), to minimize the mean square error (MSE) between \(\widetilde{g}^{\rm FDD}(\mathbf{k})\) and the ground truth \(N \tilde{f}(\mathbf{k})\), i.e.,

\begin{equation}
\begin{split}
\mathrm{MSE}(\mathbf{k}) &= \left\langle \, \left| N \tilde{f}(\mathbf{k}) 
- \tilde{g}^{\rm FDD}(\mathbf{k}) \right|^{2} \, \right\rangle\\
&\quad=\left\langle \, \left| N\tilde{f}(\mathbf{k}) 
- \sum_{l} C_l(\mathbf{k}) \tilde{g}_l(\mathbf{k}) 
\right|^{2} \, \right\rangle.
\end{split}
\label{eq:MSE}
\end{equation}

Substituting the photon density of raw FDD images $\tilde{g}_l(\mathbf{k}) 
= N\,\beta^H_l(\mathbf{k})\,\tilde{f}(\mathbf{k})+ \tilde{n}_l(\mathbf{k})$ into Eq. (\ref{eq:MSE}), we have
\begin{equation}
\begin{split}
\mathrm{MSE}(\mathbf{k})
&= \left\langle   \left|  N\tilde{f}(\mathbf{k})
    - \sum_{l} C_l(\mathbf{k})
      \big[ N\beta^H_l(\mathbf{k})\tilde{f}(\mathbf{k})  + \tilde{n}_l(\mathbf{k}) \big]\right|^{2}\right\rangle  \\
&= N^{2}
  \left|    \tilde{f}(\mathbf{k})
    - \sum_{l} C_l(\mathbf{k}) \beta^H_l(\mathbf{k}) \tilde{f}(\mathbf{k})
  \right|^{2} \\
  &+ \sum_{l}
    \left[ C_l(\mathbf{k}) \right]^{2}
    \left\langle 
      \left| \tilde{n}_l(\mathbf{k}) \right|^{2}
    \right\rangle
\end{split}
\label{eq:MSE2}
\end{equation}
The noise level $\tilde{n}_l(\mathbf{k})$ can be evaluated as follows. The mean photon number on the $i^{th}$ pixel of the detecting camera, located at $\mathbf{r}_i$, is $\langle g^{\rm DI}\left( \mathbf{r}_i \right)\rangle\Delta x^2$,  while the noise photon number is \(n_l(\mathbf{r}_i)\Delta x^2   \). For Poisson shot noise, \(\langle n_l(\mathbf{r}_i)\rangle=0\) and \(\langle n_l(\mathbf{r}_i)n_l(\mathbf{r}_{i'})\rangle=\frac{1}{\Delta x^2}\langle g_l( \mathbf{r} _i)\rangle\delta _{ii'}\). In the Fourier domain, \(\tilde{n}_l(\mathbf{k})=\iint{n_l(\mathbf{r})e^{-i\mathbf{k\cdot r}}d^2\mathbf{r}}\), thus \(\langle \tilde{n}_l(\mathbf{k})\rangle=0\) and 
\begin{equation}
\begin{split}
    \langle \left|\tilde{n}_l(\mathbf{k})\right|^2\rangle&=\iint\iint{\langle n_l(\mathbf{r})n_l(\mathbf{r'})\rangle e^{-i\mathbf{k\cdot (r-r')}}d^2\mathbf{r}d^2\mathbf{r'}}\\
    &=\sum_{i,i'}\langle n_l(\mathbf{r}_i)n_l(\mathbf{r}_{i'})\rangle e^{-i\mathbf{k\cdot (r_i-r_{i'})}}\Delta x^4\\
    &=\sum_{i,i'}\langle g_l( \mathbf{r} _i)\rangle\delta _{ii'} e^{-i\mathbf{k\cdot (r_i-r_{i'})}}\Delta x^2\\
    &=\iint{\langle g_l( \mathbf{r} )\rangle d^2\mathbf{r}}=N_l,
\end{split}
\label{eq:noise}
\end{equation}
where $N_l = N\,\beta^H_l(0) \approx \tilde{g}_l(0)$ is the total photon number recorded in the \(l^{th}\) image. Then, $\frac{\partial \mathrm{MSE}(\mathbf{k})}{\partial C_l(\mathbf{k})} = 0$ leads to 
\begin{equation}
    N^2\frac{\partial }{\partial C_l(\mathbf{k})}\left|
    \tilde{f}(\mathbf{k}) - \sum_{m} C_m(\mathbf{k}) \beta^H_m(\mathbf{k}) \tilde{f}(\mathbf{k}) \right|^{2} +2C_l(\mathbf{k})\tilde{g}_l(0) =0.
\end{equation}
Noticing that \(\beta_l\) is real, 
and assuming \(C_l\) is real, 
\begin{equation}
\begin{split}
&N^2\left[-2\left[
    \tilde{f}(\mathbf{k}) - \sum_{m} C_m(\mathbf{k}) \beta^H_m(\mathbf{k}) \tilde{f}(\mathbf{k}) \right]\beta^H_l(\mathbf{k}) \tilde{f}^*(\mathbf{k})\right] \\
    &+2C_l(\mathbf{k})\tilde{g}_l(0) =0.
\end{split}
\end{equation}
Thus,
\begin{equation}
    N^2\left| \tilde{f}(\mathbf{k})\right|^2\beta^H_l(\mathbf{k})\left[
    1 - \sum_{m} C_m(\mathbf{k}) \beta^H_m(\mathbf{k})  \right] -C_l(\mathbf{k})\tilde{g}_l(0) =0.
\end{equation}
By rearranging the terms, we obtain a system of linear equations 
\begin{equation}
A_l C_l(\mathbf{k}) + \sum_{m} \beta^H_m(\mathbf{k}) C_m(\mathbf{k}) = 1, 
\quad (l=0,1,\dots,5)
\label{eq:MSE3}
\end{equation}
where $A_l = \frac{\tilde{g}_l(0)}{\,N^2 \, |\tilde{f}(\mathbf{k})|^2 \, \beta^H_l(\mathbf{k})\,}$. The solution of Eq. (\ref{eq:MSE3}) is 
\begin{equation}
C_l(\mathbf{k})=\frac{{\beta^H_l(\mathbf{k})}/{\beta^H_l(0)}}{\sum_{m=0}^{5}{{\left[\beta^H_m(\mathbf{k})\right]^2}/{\beta^H_m(0)}}+\frac{1}{N\left|\widetilde{f}(\mathbf{k}) \right|^{2}}}, \label{eq:FDD_image_b1}     
\end{equation}
which is consistent with Eq.~(\ref{eq:FDD_image}) in the main text. For DI dcv, assuming $\tilde{g}^{\rm DI\,dcv}(\mathbf{k}) = C^{\rm DI}(\mathbf{k}) \, \tilde{g}^{\rm DI}(\mathbf{k})$ and following the same procedure from Eq. (\ref{eq:MSE}) to Eq. (\ref{eq:FDD_image_b1} ), Eq.~(\ref{eq:DI_image}) can be derived, which is just a conventional Wiener filter \cite{Ref22}.

\subsubsection{Estimation accuracy saturates CRB}
Here we prove that the unbiased estimator constructed from the reconstructed image $\tilde{g}^{\rm FDD}(\mathbf{k})$ saturates CRB of hybrid FDD.

Firstly, an estimator of the parameters $a_{\mathbf{k}}$ and ${b_{\mathbf{k}}}$ should be constructed. Note that 
\begin{equation}
\begin{split}
\tilde{f}(\mathbf{k})+\tilde{f}(\mathbf{-k})&=2\iint{f(\mathbf{r})\cos(\mathbf{k\cdot r})}d^2\mathbf{r}\\
&=2a_{\mathbf{k}}\iint{\cos^2(\mathbf{k\cdot r})}d^2\mathbf{r}=\frac{a_{\mathbf{k}}}{a_0}.
\end{split}
\end{equation}
 In the last step, we have used \(\iint{\cos^2\left( \mathbf{k \cdot r} \right)d^2\mathbf{r}}=\frac{A}{2}=\frac{1}{2a_0}\), which follows from the normalization condition of $f(\mathbf{r})$. Similarly, we have $i\tilde{f}(\mathbf{k})-i\tilde{f}(\mathbf{-k})=\frac{b_{\mathbf{k}}}{a_0}$. Since the reconstructed image $\tilde{g}^{\rm FDD}(\mathbf{k})$ is an unbiased estimator for the ground truth $N\tilde{f}(\mathbf{k})$, in the limit of \(\epsilon_{\rm FDD}(\mathbf{k})\rightarrow 0\), an unbiased estimator for  $a_{\mathbf{k}}$ can be constructed as
 \begin{equation}
     \hat a_{\mathbf{k}}=\frac{a_0}{N}\left[\tilde{g}^{\rm FDD}(\mathbf{k})+\tilde{g}^{\rm FDD}(\mathbf{-k})\right].
     \label{eq:estimator_ab}
 \end{equation}
It then  follows from Eq.~(\ref{eq:FDD_image_a})  that $\hat a_{\mathbf{k}}=\frac{a_0}{N}\sum_l C_l\left[\tilde{g}_l(\mathbf{k})+\tilde{g}_l(\mathbf{-k})\right]$. Thus, the variance of this estimator can be calculated as 

\begin{equation}
    \left< \Delta a_{\mathbf{k}}^2\right>=\frac{a_0^2}{N^2}\sum_l C_l^2\left[\langle \left|\tilde{n}_l(\mathbf{k})\right|^2\rangle+\langle \left|\tilde{n}_l(\mathbf{-k})\right|^2\rangle\right]
    \label{eq:variance}
\end{equation}
where the noise term $\langle \left|\tilde{n}_l(\mathbf{\pm k})\right|^2\rangle= N\,\beta^H_l(0)$ is derived in Eq. (\ref{eq:noise}). Substituting the expression of $C_l$ in Eq. (\ref{eq:FDD_image_b}) here, we find that in the limit of \(\epsilon_{\rm FDD}(\mathbf{k})\rightarrow 0\),

\begin{equation}
\begin{split}
    \left< \Delta a_{\mathbf{k}}^2\right>&=\frac{2a_0^2}{N}\sum_l \left[ \frac{\beta^H_l(\mathbf{k})/\beta^H_l(0)}{\sum_{m=0}^{5}{{\left[\beta^H_m(\mathbf{k})\right]^2}/{\beta^H_m(0)}}}\right]^2\beta^H_l(0)\\
    &=\frac{2a_0^2}{N}\left[ \sum_{m=0}^{5}\frac{\left[\beta^H_m(\mathbf{k})\right]^2}{{\beta^H_m(0)}}\right]^{-1}
\end{split}
    \label{eq:variance2}
\end{equation}
Comparing with Eqs.~(\ref{eq:FI_l})--(\ref{eq:FI_hybrid}), we see that $\left< \Delta a_{\mathbf{k}}^2\right>=\frac{1}{N}\left({\rm CRB}_{\rm hybrid}^{\rm FDD}\right)_{a_{\mathbf{k}},a_{\mathbf{k}}}$, i.e., this estimator saturates CRB. Similar analysis works for $b_{\mathbf{k}}$ as well.

\end{document}